\documentclass[iop]{emulateapj}

\usepackage{graphicx}
\usepackage{amsmath}
\usepackage{subfigure}
\usepackage{float}
\usepackage{natbib} 
\usepackage{hyperref}
\usepackage{tabularx}

\begin{document}

\title{EvryFlare III: Temperature Evolution and Habitability Impacts of Dozens of Superflares Observed Simultaneously by Evryscope and TESS}

\author{Ward S. Howard\altaffilmark{1}, Hank Corbett\altaffilmark{1}, Nicholas M. Law\altaffilmark{1}, Jeffrey K. Ratzloff\altaffilmark{1}, Nathan Galliher\altaffilmark{1}, Amy L. Glazier\altaffilmark{1},  Ramses Gonzalez\altaffilmark{1}, Alan Vasquez Soto\altaffilmark{1}, Octavi Fors\altaffilmark{1,2}, Daniel del Ser\altaffilmark{1,2}, Joshua Haislip\altaffilmark{1}}
\altaffiltext{1}{Department of Physics and Astronomy, University of North Carolina at Chapel Hill, Chapel Hill, NC 27599-3255, USA}
\altaffiltext{2}{Dept. de F\'{\i}sica Qu\`antica i Astrof\'{\i}sica, Institut de Ci\`encies del Cosmos (ICCUB), Universitat de Barcelona, IEEC-UB, Mart\'{\i} i Franqu\`es 1, E08028 Barcelona, Spain}

\email[$\star$~E-mail:~]{wshoward@unc.edu}

\begin{abstract}
Superflares may provide the dominant source of biologically-relevant UV radiation to rocky habitable-zone M-dwarf planets (M-Earths), altering planetary atmospheres and conditions for surface life. The combined line and continuum flare emission has usually been approximated by a 9000 K blackbody. If superflares are hotter, then the UV emission may be 10$\times$ higher than predicted from the optical. However, it is unknown for how long M-dwarf superflares reach temperatures above 9000 K. Only a handful of M-dwarf superflares have been recorded with multi-wavelength high-cadence observations. We double the total number of events in the literature using simultaneous Evryscope and TESS observations to provide the first systematic exploration of the temperature evolution of M-dwarf superflares. We also increase the number of superflaring M-dwarfs with published time-resolved blackbody evolution by $\sim$10$\times$. We measure temperatures at 2 min cadence for 42 superflares from 27 K5-M5 dwarfs. We find superflare peak temperatures (defined as the mean of temperatures corresponding to flare FWHM) increase with flare energy and impulse. We find the amount of time flares emit at temperatures above 14,000 K depends on energy. We discover 43\% of the flares emit above 14,000 K, 23\% emit above 20,000 K and 5\% emit above 30,000 K. The largest and hottest flare briefly reached 42,000 K. Some do not reach 14,000 K. During superflares, we estimate M-Earths orbiting $<$200 Myr stars typically receive a top-of-atmosphere UV-C flux of $\sim$120 W m$^{-2}$ and up to 10$^3$ W m$^{-2}$, 100-1000$\times$ the time-averaged XUV flux from Proxima Cen.
\end{abstract}

\keywords{Exoplanet atmospheres, Ultraviolet astronomy, Astrobiology, Stellar flares, Optical flares}

\maketitle

\section{Introduction}

Stellar flares are stochastic events that occur when a star's magnetic field re-connects, releasing intense radiation across the electromagnetic spectrum \citep{Kowalski2013}. Rocky planets in the habitable zones (HZ) of M-dwarfs (M-Earths) are often subjected to superflares \citep{tarter2007,Howard2019,Gunter2020}, flare events with energy $\geq$10$^{33}$ erg and 10-1000$\times$ the energy of the largest solar flares \citep{Schaefer2000}. Frequent superflares can erode the ozone layer of an Earth-like atmosphere and allow lethal amounts of UV surface flux \citep{Segura2010, Howard2018, Tilley2019}. Conversely, too few flares may result in insufficient surface radiation to power pre-biotic chemistry due to the inherent faintness of M-dwarfs in the UV \citep{Ranjan2017,Rimmer2018}. 

Superflares are rare \citep{Lacy1976}, making observations difficult. Drawing broad conclusions about the temperatures of superflares or their impacts on exoplanets is stymied by only a handful of observations. For example, it is currently unknown whether the thermal emission of superflares is consistently higher than for typical flares. It is also unknown if the impulse (i.e. how peaked a flare appears in photometry) is consistently higher for hot superflares. Errors in the temperatures of optical superflares propagate to the estimated UV emission that determines the space weather environment.

Flares radiate energy in both emission lines and in the continuum, with the continuum dominating the energy budget from the FUV to the optical. During the peak phase of most flares, only $\sim$4\% of the total energy is found in the emission lines. During the gradual decay phase, line emission may contribute 20\% of the total energy budget \citep{Kowalski2013}. In several flares, however, line emission has been found to contribute up to 50\% of the total flare energy \citep{Hawley2007}.

The combined line and continuum emission of stellar flares has often been approximated by a 9,000-10,000 K blackbody \citep{Osten2015}. The blackbody temperature governs the energy budget of the flare, especially the fraction of the energy emitted at the UV wavelengths that most strongly react with exoplanet atmospheres \citep{Kowalski2018}. The canonical value of 9000 K provides a lower limit to the energy emitted in the UV, with higher-temperatures resulting in more UV radiation. The effective blackbody temperatures of superflares are tremendously uncertain. Continuum temperatures of M-dwarf flares usually range from 9000 K to 14,000 K \citep{Kowalski2013} but temperatures may extend above 40,000 K \citep{Robinson2005, Kowalski_Allred2018, Froning2019}. Significant temperature changes occur over the course of individual flares as the dominant source of flare heating transitions from the base of the stellar atmosphere into the corona \citep{Kowalski2013}. The blackbody temperature is a key ingredient in modeling the effects of optical superflares upon the atmospheric photochemistry of Earth-like planets. The UV energy of a $\sim$30,000 K optical superflare computed assuming a 10,000 K blackbody will be under-estimated by a factor of 16 \citep{Planck1901}. Furthermore, temperatures in the FUV in excess of 40,000 K increase the rate of photo-dissociation in exoplanet atmospheres by 10-100$\times$ \citep{Loyd2018, Froning2019}.

Few M-dwarf superflares have been observed with UV colors directly. Two examples of such events include the Great Flare of AD Leo \citep{Hawley1991} and the Hazflare \citep{Loyd2018hazflare}. The Great Flare of AD Leo (M4) released 10$^{34}$ erg and exhibited a continuum consistent with a temperature of 9000 K. To date, most atmospheric modeling of potentially-habitable planets orbiting flare stars assume spectral evolution templates based upon this singular event \citep{Segura2010, Howard2018, Loyd2018hazflare, Tilley2019}. The Hazflare (emitted by a 40 Myr M2 dwarf) released 10$^{33.6}$ erg with a blackbody temperature of 15,500 K. Large flares observed by the Galaxy Evolution Explorer (GALEX; \citealt{Bianchi1999}) in the UV have reached temperatures as high as 50,000 K, measured from FUV and NUV colors. Most superflares from cool stars seen by GALEX are from late K-dwarfs, with only 1 superflare recorded from an M-dwarf in \citet{Robinson2005} and 7 superflares from 4 M-dwarfs recorded in \citet{Brasseur2019}.

Multi-wavelength superflares in other band-passes indirectly estimate UV emission through the blackbody. However, non-thermal emission may lead to under-estimates of the UV emission. Multi-wavelength superflare observations are uncommon. Apart from the GALEX events, only 19 superflares from 13 M-dwarfs have been recorded with multi-wavelength, high cadence observations since 1976 \citep{Lacy1976, Doyle1988, Hawley1991, Hawley1995, Garcia_Alvarez2002, Kowalski2010, Osten2010, Osten2016, Loyd2018hazflare, Luo2019, Namekata2020} \footnote{The count includes only flares recorded on ADS with clearly-quoted integrated flare energies. Events were required to have $\sim$2 min or higher cadence with simultaneous multi-wavelength observations or spectra.}. These known flare stars include AD Leo, YZ CMi, EQ Peg, EV Lac, UV Ceti, CN Leo, Wolf 424 AB, YY Gem, GL 644 AB, AT Mic, DG Cvn, the Tuc Hor star GSC 8056-0482, and BX Tri. These stars were largely selected for monitoring during ``staring" campaigns designed to capture stochastic flaring, biasing the sample toward a handful of extremely active stars.

We measure blackbody temperatures of dozens of superflares observed simultaneously at 2 min cadence by the Evryscope \citep{Law2016,Ratzloff2019} and Transiting Exoplanet Survey Satellite (TESS; \citealt{Ricker_TESS}) surveys as shown in Figure \ref{fig:grid_evr_vs_TESS_flares}. Each Evryscope is an array of small telescopes imaging the entire accessible sky all at once. Each night, Evryscope-South observed the entire TESS Cycle 1 field simultaneously in $g^{\prime}$-band for hours. In order to obtain a representative sample of superflares from stars of various activity levels, we search multi-band photometry of hundreds of late-type stars. With the 2 min cadence of Evryscope and TESS, we robustly quantify the amount of time superflares emit at temperatures in excess of 9000 K.

In Section \ref{EvryFlare}, we describe the simultaneous flare observations and host star properties. In Section \ref{tess_flares}, we describe our flare search methods. In Section \ref{flare_energies}, we describe how flare energies are computed in each bandpass. In Section \ref{measuring_BB}, we describe how effective blackbody temperatures are computed for each flare. In Section \ref{results}, we describe how flare energy and morphology predict temperature. In Section \ref{hot_flares}, we describe how stellar mass and age impact the UV-C space weather environments of orbiting terrestrial planets. In Section \ref{conclude}, we conclude.

\section{Photometry}\label{EvryFlare}
We discover flares in photometry from TESS and Evryscope.

\subsection{Evryscope observations}\label{evryscope_observations}
Evryscope-South is located at Cerro Tololo Inter-American Observatory in Chile, and Evryscope-North is located at Mount Laguna Observatory in California, USA. Each Evryscope is an all sky array of small telescopes that continuously and simultaneously images 8150 square degrees and 18,400 square degrees in total each night at a resolution of 13\arcsec pixel$^{-1}$ and down to an airmass of two. Evryscope-South observes at two-minute cadence in \textit{g}\textsuperscript{$\prime$}~\citep{Evryscope2015} for $\sim$6 hr each night and has a typical dark-sky limiting magnitude of \textit{g}\textsuperscript{$\prime$}=16 \citep{Ratzloff2019}. The system accomplishes this coverage by employing a ``ratchet" strategy that tracks the sky for 2 hours before ratcheting back into the initial position and continuing observations.

Evryscope images are processed using the Evryscope Fast Transient Engine (EFTE; \citealt{Corbett2020AAS}). EFTE performs simple aperture photometry at the catalog location of each source. The resulting magnitudes are calibrated to the ATLAS All-Sky Stellar Reference Catalog \citep{Tonry2018} using a smoothly interpolated zero-point across each individual camera's field of view. 

\subsection{TESS observations}\label{tess_observations}
The TESS mission is looking for transiting exoplanets across the entire sky, split into 26 sectors. TESS observes each sector continuously with four 10.5 cm optical telescopes in a red (600-1000 nm) bandpass for 28 days at 21$\arcsec$ pixel$^{-1}$. Calibrated, short-cadence TESS light curves of each flare star were downloaded from MAST\footnote{https://mast.stsci.edu}. We selected Simple Aperture Photometry (SAP) light curves rather than Pre-search Data Conditioning (PDC) ones to avoid altering the impulsive flare structure.

\begin{figure*}
	\centering
	{
		\includegraphics[trim= 1 1 1 1,clip, width=6.9in]{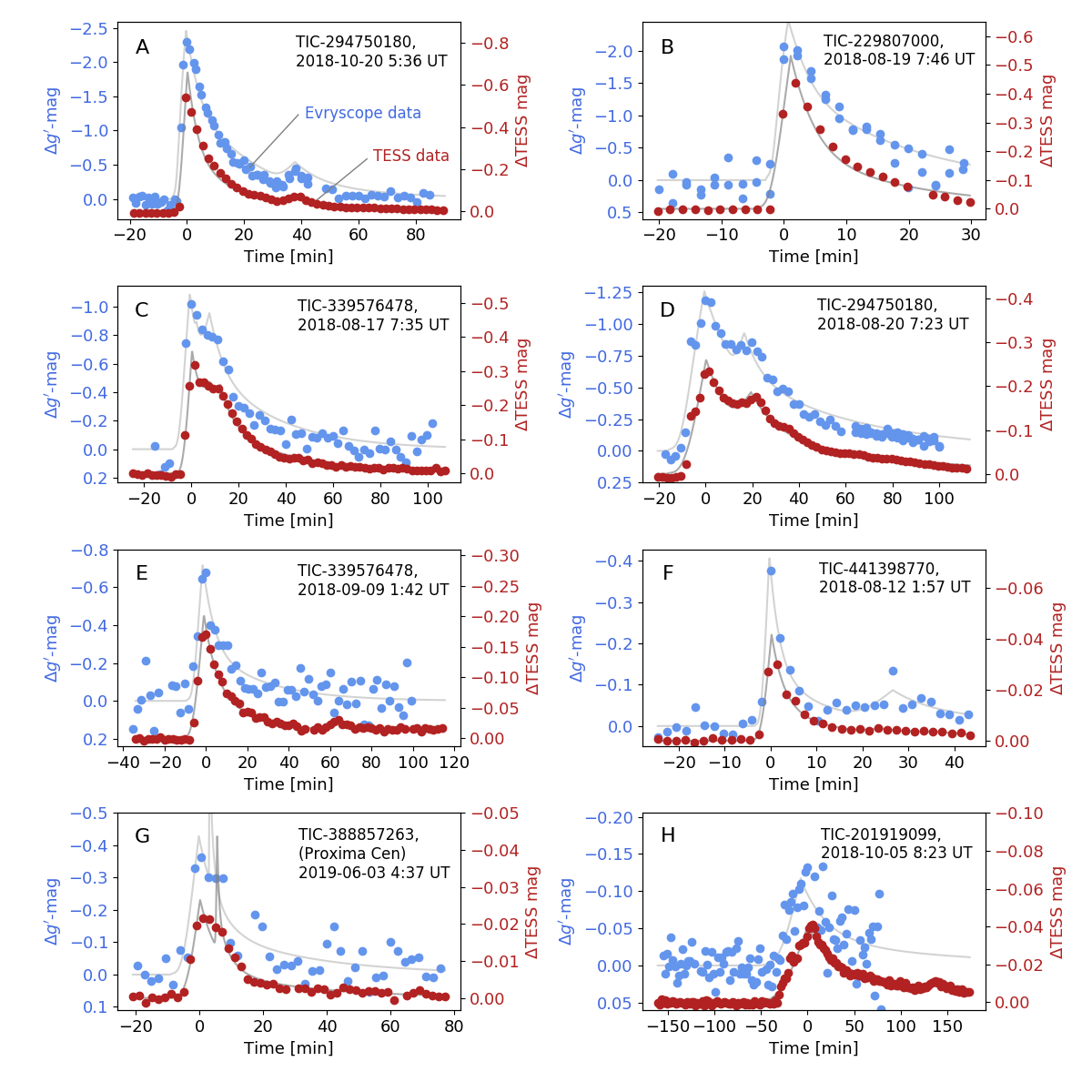}
	}
	\caption{Simultaneous flare events observed at 2 min cadence by Evryscope (blue) and TESS (red) for multi-wavelength coverage. Multi-wavelength, high-cadence observations of superflares are necessary to understand their influence on the evolution of planet atmospheres; however such observations have been previously obtained for only a handful of superflares from a handful of highly-active stars such as AD Leo. Our sample of 44 large flares expands our knowledge to a diverse population of cool dwarfs. Flare fits are shown as grey lines. UT identifiers are approximated from barycentric TESS epochs and may differ by up to 10 min from the exact flare peak time.}
	\label{fig:grid_evr_vs_TESS_flares}
\end{figure*}

\begin{table*}
\renewcommand{\arraystretch}{1.6}
\caption{Temperatures of Simultaneous Flares Observed During TESS Cycle 1}
\begin{tabular}{p{1.3cm} p{2.2cm} p{0.5cm} p{0.5cm} p{0.7cm} p{0.7cm} p{1.3cm} p{1.5cm} p{1.5cm} p{0.4cm} p{1.2cm} p{0.8cm} p{0.7cm}}
\hline
 &  &  &  &  &  &  &  &  &  &  &  & \\
TIC-ID & Obs. & log E$_{g^{\prime}}$ & log E$_T$ & A$_{g^{\prime}}$ & A$_T$ & Color & Peak T$_\mathrm{Eff}$ & Tot. T$_\mathrm{Eff}$ & t$_\mathrm{Abv}$ & Impulse & P$_\mathrm{Rot}$ & M \\
 & [UT] & [Erg] & [Erg] & [$\Delta$F/F] & [$\Delta$F/F] & [A$_g$-A$_T$] & [K] & [K] & [Min] & [A$_g$/Min] & [d] & [M$_\odot$] \\
\hline
 &  &  &  &  &  & ... &  &  &  &  &  & \\
294750180 & 2018-10-20 05:36 & 34.7 & 34.4 & 7.24 & 0.65 & 6.59$\pm$0.04 & 34000$\pm$2300 & 18600$\substack{+600 \\ -600}$ & 12.6 & 1.3$\pm$0.2 & 0.5255 & 0.54 \\
229807000 & 2018-08-19 07:46 & 34.3 & 34.2 & 5.76 & 0.5 & 5.26$\pm$0.13 & 15500$\pm$400 & 12100$\substack{+700 \\ -600}$ & 2.9 & 0.9$\pm$0.1 & 0.3746 & 0.36 \\
382043650 & 2018-11-09 05:49 & 34.0 & 34.0 & 1.68 & 0.3 & 1.38$\pm$0.12 & 15100 $\pm$300 & 7500$\substack{+500 \\ -500}$ & 0.0 & 0.14$\pm$0.02 & 6.13 & 0.29 \\
5796048 & 2018-09-08 03:42 & 33.8 & 33.9 & 2.34 & 0.26 & 2.08$\pm$0.13 & 6900 $\pm$300 & 6900$\substack{+700 \\ -700}$ & 0.0 & 0.45$\pm$0.09 & 0.5557 & 0.42 \\
339576478 & 2018-08-17 07:35 & 35.0 & 34.9 & 1.56 & 0.34 & 1.22$\pm$0.12 & 18400$\pm$600 & 11800$\substack{+2100 \\ -1700}$ & 6.0 & 0.1$\pm$0.01 & 2.113 & 0.57 \\
 &  &  &  &  &  & ... &  &  &  &  &  & \\
\hline
\end{tabular}
\label{table:indiv_flares_tab}
{\newline\newline \textbf{Notes.} A subset of the parameters of 44 large flares observed simultaneously by Evryscope and TESS. The full table is available in machine-readable form. Columns are TIC-ID, observation date and time, Evryscope flare energy, TESS flare energy, Evryscope amplitude in normalized flux, TESS amplitude in normalized flux, flare color between normalized flux amplitudes, peak temperature (ave. temperature in FWHM), total temperature, time spent above 14,000 K, impulse, stellar rotation period, and stellar mass. UT observation identifiers are approximated from barycentric TESS epochs and may differ by up to 10 min from the exact flare peak time.}
\end{table*}

\subsection{The EvryFlare stellar sample}\label{flare_star_sample}

The EvryFlare survey is an ongoing search for stellar activity from all cool stars observed by the Evryscopes across the accessible sky. We select 306 K5-M6 flare stars observed simultaneously by Evryscope and TESS during Cycle 1. The EvryFlare targets are some of the nearest (23$\substack{+27 \\ -12}$ pc) and brightest ($g^{\prime}$=12.3$\substack{+1.4 \\ -1.0}$) flare stars. A number of flare stars in our sample are also planet-search targets, with 3 targets (Proxima Cen, LTT 1445, and the WD+M4 binary RR Cae (AB)) already known to host planets; constraining their space weather will benefit future atmospheric characterization \citep{Howard2019}.

We obtain spectral types of our late K and M-dwarf stars from SIMBAD \citep{Wenger2000} if possible and then from \citet{Howard2019b}. SIMBAD types were used for 80\% of our flares, and the analysis from \citet{Howard2019b} incorporating multi-band photometry and stellar distances was used for the other 20\%. We use SIMBAD if available because nearly all (97\%) of the SIMBAD entries of flares in our sample had types classified by spectra, and spectral absorption lines in M-dwarf reconnaissance spectra are a reliable way to identify sub-types in our mass range. Most of the SIMBAD types were identified by several spectroscopic surveys (e.g. \citealt{2002AJ....123.2002H,2006AJ....132.2360H, 2006AJ....132..866R, 2006A&A...460..695T,2014AJ....147..146K,2014MNRAS.443.2561G,2017ApJ...840...87R}). The stellar mass is estimated from the spectral type using \citet{Kraus2007}. Most stellar rotation periods are from \citet{Howard2019b}; missing periods are supplemented by strong rotational modulation seen in the TESS light curves.

\newpage

\section{Identifying simultaneous TESS and Evryscope flares}\label{tess_flares}
We identify flares in the TESS light curves, then search the Evryscope light curves for simultaneous events. We pre-whiten the TESS light curves of sinusoidal variability and systematics before searching for flares. We then search each TESS light curve for large flares by selecting photometric brightening events 5$\sigma$ above the local photometric noise. Flares in TESS smaller than 5$\sigma$ above the noise are not considered since they will likely be too noisy for high-signal detections in the lower-precision Evryscope light curves. We remove any TESS flare candidates that occur between 9:00 and 23:00 UT. We vet each remaining candidate in the TESS light curves by eye, removing common TESS systematics and marginal detections. We consider multiple flare peaks within the start and stop time to be a single event. In sum, we find 806 TESS flares that occur during the night from 163 K5-M5 stars. We search the Evryscope database at the coordinates and time of each flare event and produce light curves where we have data. We visually inspect each Evryscope light curve at the time of the TESS flare, looking for epochs exhibiting rapid-rise, slow-decay profiles that exceed the local noise. Because low-signal events do not produce useful data for temperature measurement, we only include flares large enough to be clearly visible by eye and that produce a well-defined flux increase during the rapid phase.

We convert the MJD time stamps of the Evryscope light curves into barycentric MJD (BMJD). Because TESS time stamps are recorded in the middle of each exposure while Evryscope stamps are recorded at the beginning, we must add 1 min to correct offsets between the surveys. Next, we remove any flares with peak amplitudes comparable to the noise in the Evryscope light curve ensuring clear signals for flare temperature measurement. 

\begin{figure*}
	\centering
	{
		\includegraphics[trim= 1 1 1 1,clip, width=6.9in]{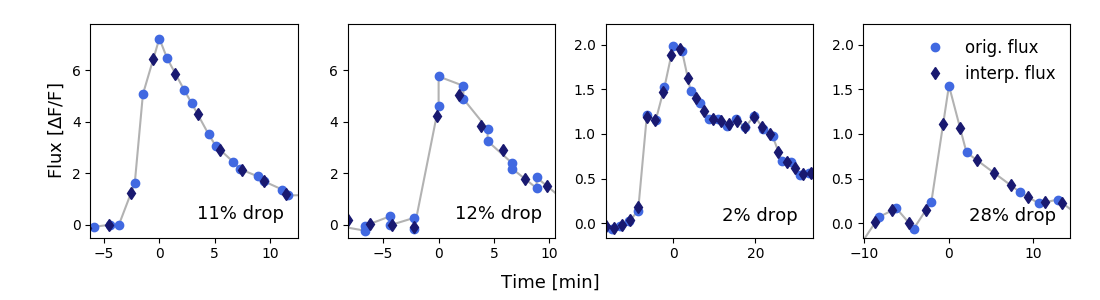}
	}
	\caption{Evryscope light curves of several of the largest flares are shown in light blue. In dark blue, the epochs are interpolated to the timestamps of the TESS epochs to provide simultaneous flux estimates. While the peak epoch of each flare is under-estimated at 2 min cadence by the linear interpolation process, it is necessary to sync up the physics occurring in each light curve: the two closest epochs in Evryscope and TESS can occur up to 1 min apart, but impulsive flare heating events may evolve at timescales much faster than 1 min. Ultimately, the new TESS 20 s cadence mode will alleviate the need for this technique.}
	\label{fig:interpolated_flares}
\end{figure*}

To place both surveys on an identical time axis, linear interpolation of the Evryscope epochs, fluxes, and uncertainties is performed, and subsequently evaluated at the TESS timestamps. Because impulsive flare heating features in optical light curves may evolve at timescales as short as $\sim$10 s \citep{Moffett1974}, it would be best to observe each flare at $\sim$10 s cadence with TESS. At 2 min cadence however, we may either (1) directly compare flux values at the Evryscope and TESS epochs with the smallest separation in time, which assumes the physics captured by timestamps up to 1 min apart is still simultaneous, or (2) estimate simultaneous behavior using linear interpolation for timestamps within each two minute window. Option (1) biases instantaneous temperatures since different impulsive flare heating events may be recorded even 1 minute apart. We choose option (2) since it attempts to minimize the time lag between the physics captured in each light curve. This process does not significantly alter the profile of most superflares, which typically last for tens of minutes to hours; it may under-represent the rapidly-changing flux near the flare peak resulting in a lower temperature. The drop in flux depends on how far away the Evryscope and TESS timestamps are from each other. Essentially no under-estimation of the peak occurs when the timestamps are within seconds of one another; $\sim$25\% percent drops are possible when timestamps differ by up to 1 min as shown in Figure \ref{fig:interpolated_flares}. The peaks of the largest flares (i.e. those that increase the stellar brightness by a factor of 2) are under-estimated by a median and 1$\sigma$ drop of 8$\substack{+18 \\ -8}$\%. The peaks of the smaller flares are under-estimated by a median and 1$\sigma$ drop of 5$\substack{+14 \\ -4}$\%. All simultaneous flares detected in this survey, their temperatures, and their host star information are available in Table \ref{table:indiv_flares_tab}.

\section{Flare energetics}\label{flare_energies}
We measure the quiescent luminosity of each flare star in erg $s^{-1}$ for the $g^{\prime}$ and TESS bandpasses, respectively. The quiescent luminosity is computed using the method of \citet{Howard2018, Howard2019}, which relies upon the $g^{\prime}$=0 to flux calibration \citep{Hewett2006}, the $T$=0 to flux calibration \citep{Sullivan2015}, the stellar distance, the $g^{\prime}$ magnitude, and the $T$ magnitude of the star. Stellar distances and uncertainties are primarily obtained from the TESS Input Catalog version 8 (TIC; \citealt{Bailer_Jones2018,Stassun2019}). $g^{\prime}$ magnitudes and uncertainties are primarily obtained from the Asteroid Terrestrial-impact Last Alert System (ATLAS) catalog \citet{Tonry2018}. TESS magnitudes and uncertainties are obtained from the TIC. Errors in distance and apparent magnitude in each band-pass are propagated to the quiescent luminosity. The quiescent luminosity of 8 stars found in stellar binaries is corrected in both Evryscope and TESS data for blending due to the brightest two stars within 21" of each target pixel, preventing significant under-estimates of the flare energy in most cases.

Flare start and stop times are determined from the TESS light curves as the initial and final epochs near the flare peak that exceed 1$\sigma$ above the photometric noise. TESS start and stop times are subsequently adjusted by eye to include the flare tail. Start and stop times are used only for the purposes of providing limits of integration, so we do not worry about the 1 min before the rapid rise is timestamped at the middle of the 2 min exposure- this flux is included in the integration and not lost. The fractional flux is calculated as described in \citet{hawley2014}. Fractional flux is computed as $\Delta$F/F=$\frac{\mid F-F_0 \mid}{F_0}$ where $F_0$ is the out-of-flare flux and is determined from the local median. The uncertainty in F$_0$ is determined by a bootstrap analysis of the median out-of-flare flux within a few hours of each event. The equivalent duration (ED) for the entire duration of each flare is calculated as described in \citet{hawley2014}, between upper and lower limits of the flare start and stop times. The energies computed for each flare within the FWHM of the flare peak are computed between the start and stop times at which the flare exceeds half its peak amplitude. The quiescent luminosity in each bandpass is multiplied by the ED in each bandpass to measure energy. Errors in energy are computed with 200 MC trials varying each input.

We fit the \citet{davenport2014} flare model to each flare's light curve. Flares are often best-fit by a superposition of multiple emission events. We visually inspect each light curve and determine the number of flare peaks in each event, then fit a superposition of 1-3 flares. Flare amplitude, FWHM, and timing are all set as free parameters. TESS and Evryscope light curves of the same events are fitted separately to allow for differences between band-passes. Looking ahead to Section \ref{model_temperatures}, the purpose of fitting the flare light curves with a model is to provide a second estimate of the flare temperature evolution; a smoothly-varying function reduces noise in the individual temperature measurements obtained at 2 min cadence during the flare decay. It is useful to compare both the measured temperature evolution and the model temperature evolution for each flare.

\begin{figure}
	\centering
	{
		\includegraphics[trim= 1 1 1 1,clip, width=3.4in]{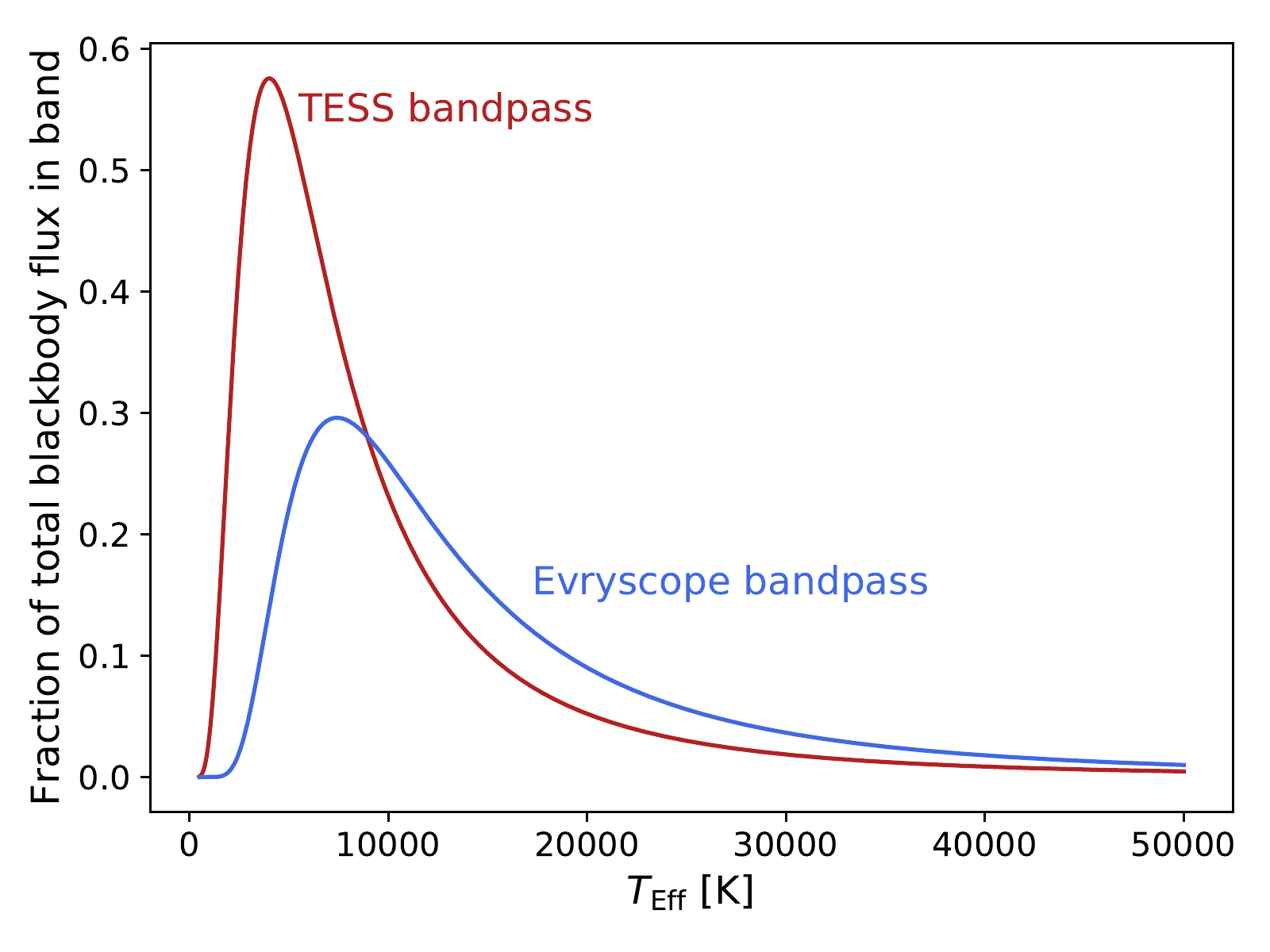}
	}
	\caption{The fraction of the total blackbody flux in the Evryscope and TESS bands is shown versus the blackbody temperature. The highest fraction of flux relative to temperature seen in the red TESS bandpass will occur for 4000 K flares, while the highest fraction of flux seen in the blue Evryscope $g^{\prime}$ bandpass will occur at 7300 K. The two curves converge above $\sim$46000 K, dis-allowing temperature estimation. This plot is inspired by Figure 12 of \citet{Schmitt2019}.}
	\label{fig:schmidt_style_figure}
\end{figure}

\begin{figure*}
	\centering
    \subfigure
	{
		\includegraphics[trim= -30 1 0 -1, clip, width=3.4in]{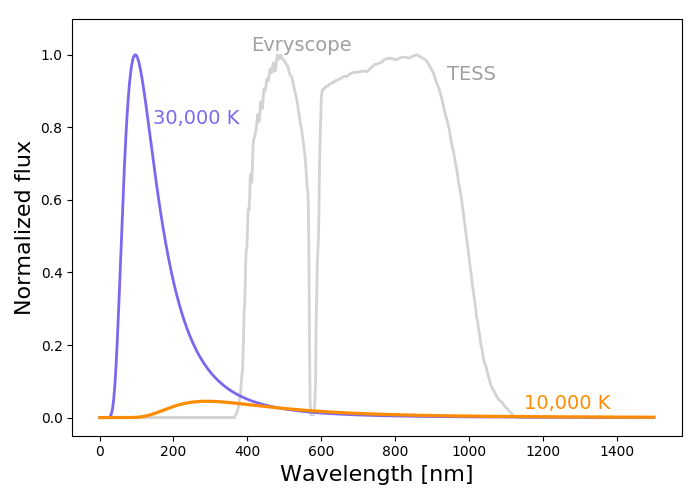}
		\label{fig:evr2}
	}
	\subfigure
	{
		\includegraphics[trim= -2 10 -2 -20, clip, width=3.4in]{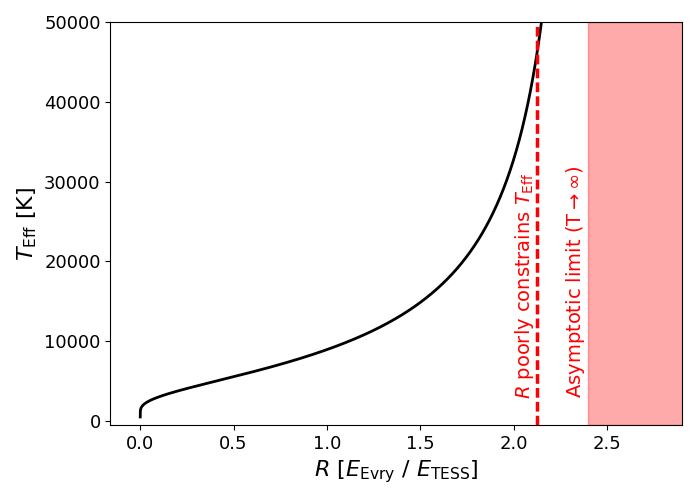}
		\label{fig:evr}
	}
	\subfigure
	{
		\includegraphics[trim= 0 1 0 1, clip, width=3.4in]{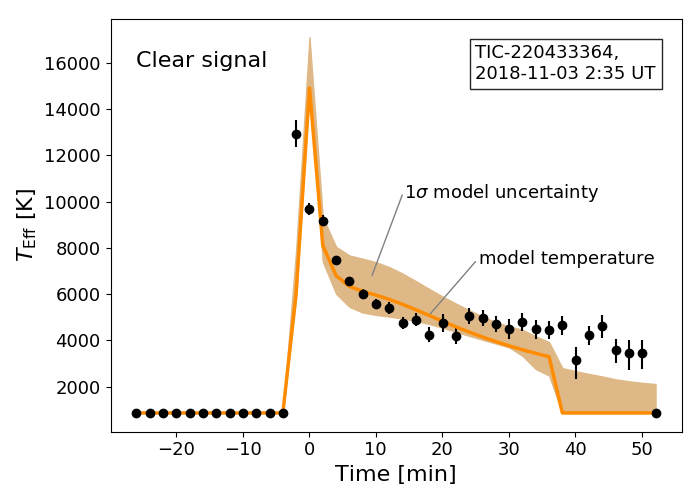}
		\label{fig:ampl2}
	}
	\subfigure
	{
		\includegraphics[trim= 0 1 0 1, clip, width=3.4in]{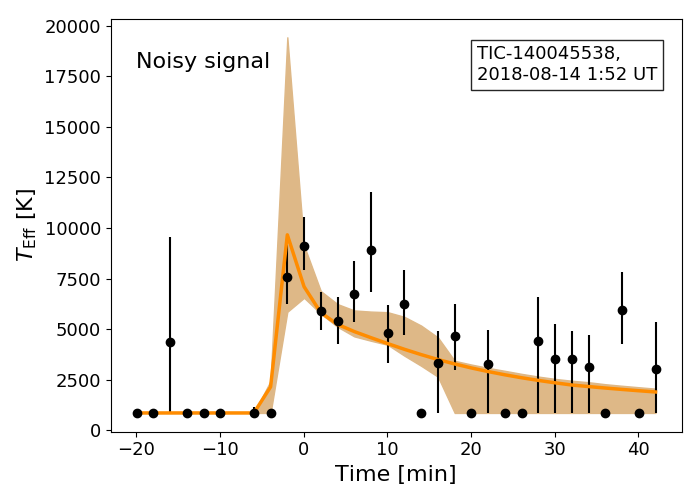}
		\label{fig:ampl}
	}
	\caption{Methods. Top left: a 10,000 K flare blackbody compared to a 30,000 K blackbody. The Evryscope and TESS band-passes are overlaid. The UV energy of a hot flare may be under-estimated from the optical by $\sim$10$\times$ if the canonical temperature is assumed. Top right: The ratio of flare energies observed in the Evryscope and TESS band-passes uniquely determines the effective blackbody temperature. A blackbody of temperature $T_\mathrm{Eff}$ is separately convolved with the response functions of each instrument to produce a ratio $R$. The wide wavelength range of TESS offsets reduced emission at longer wavelengths, allowing sensitivity to temperatures as high as 46,000 K. Above this value, the $T_\mathrm{Eff}$-$R$ relation becomes asymptotic due to insufficient flux in the TESS bandpass. Bottom panels: temperatures of individual epochs are compared with temperatures from the flare models fitted to the light curves. We confirm our model temperatures broadly reproduce trends in the data for low-uncertainty flare signals (left) and use our models to understand the temperature evolution of high-uncertainty data (right).}
	\label{fig:temperature_methods}
\end{figure*}

\begin{figure*}
	\centering
	{
		\includegraphics[trim= 1 1 1 1,clip, width=6.9in]{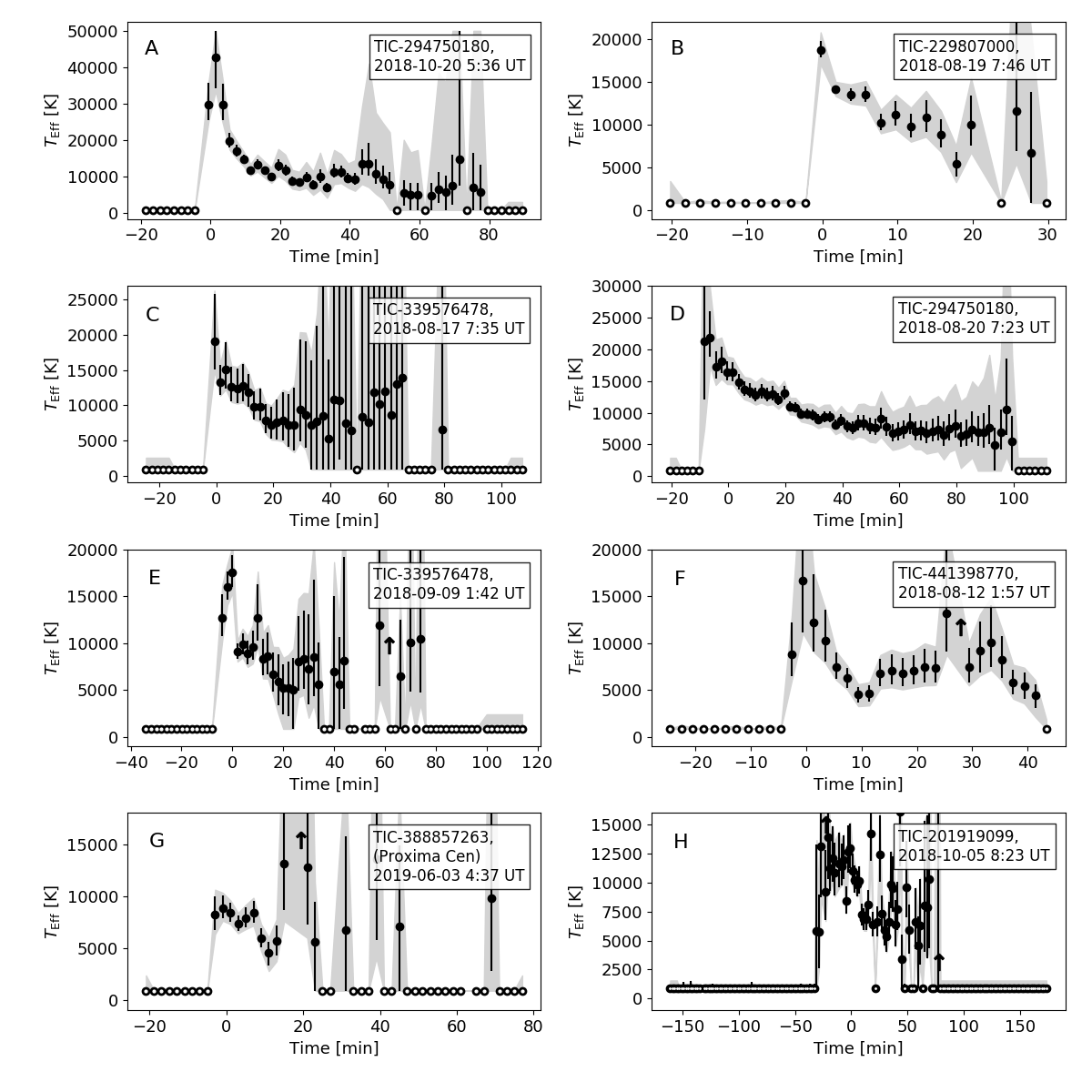}
	}
	\caption{The continuum temperature evolution of the sample flares identified in Figure \ref{fig:grid_evr_vs_TESS_flares}. Temperature measurements of 44 flares were obtained at 2 min cadence, providing a statistical sample of how long M-dwarf superflares emit at high temperatures. Flare A is our largest and hottest flare, briefly peaking at 42,000 K. Formal errors are represented in black and systematic errors in grey. Temperature non-detections are displayed as hollow circles.}
	\label{fig:grid_temperatures}
\end{figure*}

\section{Flare temperature Methods}\label{measuring_BB}

We define the color-temperature of a flare as the effective continuum temperature inferred from a flare's spectral properties. We measure the color-temperature as follows:
\begin{enumerate}
  \item We compute the radiation spectrum for a blackbody of temperature $T$ as a function of wavelength using Planck's law from 1 to 1500 nm, ensuring coverage of the tail of the Planck curve beyond the TESS bandpass.
  \item We separately convolve the blackbody spectrum with the Evryscope and TESS response functions and integrate over all wavelengths to obtain the energy in each bandpass. The fraction of total blackbody flux in each band is shown in Figure \ref{fig:schmidt_style_figure}.
  \item We take the ratio $R$ of the energy observed in the Evryscope bandpass to the energy in the TESS bandpass.
  \item We repeat the above process for blackbody temperatures from 500 K to 50,000 K to create a one-to-one function $R(T_\mathrm{Eff})$. \item Since our data is in the form of Evryscope and TESS light curves, we estimate the values of $R$ for each flare using the values of observed flare energies in the two bandpasses.
  We therefore invert the function to find $T_\mathrm{Eff}(R)$.
\end{enumerate}
The function $T_\mathrm{Eff}(R)$ is plotted in Figure \ref{fig:temperature_methods}. Because both the Evryscope and TESS bandpasses are in the tail of the Planck curve, high temperatures result in very small amounts of energy in the TESS bandpass, making a temperature determination increasingly difficult. Our $R$-$T_\mathrm{Eff}$ relationship indicates the temperature information that may be gleaned from our bandpasses ceases above $\sim$46,000 K when $T_\mathrm{Eff}(R)$ becomes asymptotic.

We note this method is essentially identical to that of Equation 1 of \citet{Hawley_2003}, except that we do not simultaneously fit the flare area since it cancels out with two colors. We reproduce their Equation 1 here. $F_\mathrm{\lambda}$ is the flux in a given bandpass, $A_\mathrm{fl}$ is the projected area of the flare, defined as $X \pi R_{*}^2$, $X$ is the flare area as a fraction of the projected stellar area $A_{*}=\pi R_{*}^2$, $d$ is the stellar distance, and $B_{\lambda}(T_\mathrm{fl})$ is the emission of a blackbody of temperature $T_\mathrm{fl}$ in a given bandpass. 
\begin{equation}
    F_\mathrm{\lambda}=\frac {A_\mathrm{fl} B_{\lambda}(T_\mathrm{fl})}{d^2}
    \label{eq:notsu_spot_flare_scaling1}
\end{equation}
Assuming $A_\mathrm{fl}$ does not vary appreciably between the $g^{\prime}$ and $T$ bandpasses and the areas cancel, dividing $F_{g^{\prime}}$ by $F_\mathrm{T}$ results in the following equation:
\begin{equation}
    \frac {F_{g^{\prime}}} {F_\mathrm{T}} = \frac {X_{g^{\prime}} B_{g^{\prime}}(T_\mathrm{fl})}{X_\mathrm{T} B_\mathrm{T}(T_\mathrm{fl})} \approx \frac {B_{g^{\prime}}(T_\mathrm{fl})}{B_\mathrm{T}(T_\mathrm{fl})}
    \label{eq:notsu_spot_flare_scaling2}
\end{equation}
This last expression is identical to the approach we used to estimate flare temperatures using only the Evryscope and TESS bandpasses. Because a flare's spatial extent should be approximately the same in either bandpass, we may solve the system of equations for $T_\mathrm{fl}$ and $A_\mathrm{fl}$ separately. More recently, \citet{Castellanos2020} make this same assumption in their Appendix A, Equation 4 to find the same result as our Equation \ref{eq:notsu_spot_flare_scaling2} if $T_\mathrm{fl}/T_*>>1$ or $T_* \approx 0$ K. They note this approximation is valid for M-dwarfs but not G-dwarf stars.

\begin{figure*}
	\centering
    \subfigure
	{
		\includegraphics[width=3.4in]{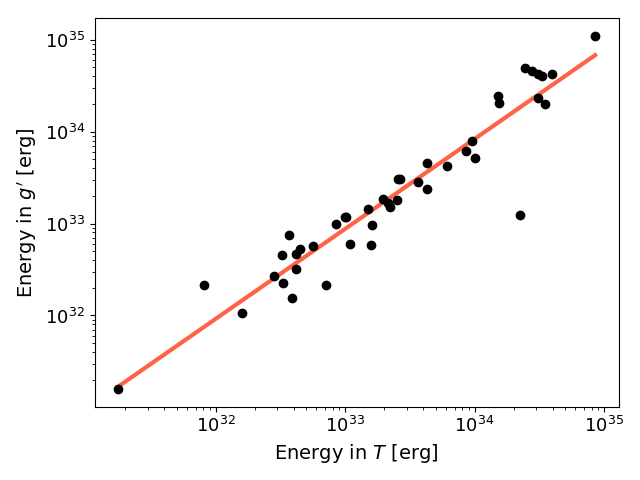}
		\label{fig:pwrlaw1}
	}
	\subfigure
	{
		\includegraphics[width=3.4in]{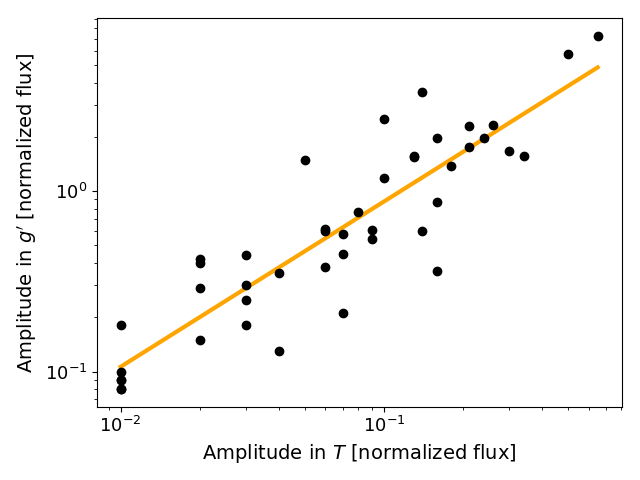}
		\label{fig:pwrlaw2}
	}
	\caption{Scaling relationships for flare energy (left) and amplitude (right) between the Evryscope and TESS bands. While Evryscope amplitudes are approximately an order of magnitude larger than the TESS amplitudes, the energies are comparable between band-passes.}
	\label{fig:evry_tess_pwrlaws}
\end{figure*}

Two min cadence allows three unique measurements of the flare temperature:
\begin{enumerate}
  \item We compute flare temperatures epoch-by-epoch across the flare light curve to understand how the temperature changes with time, demonstrated in Figure \ref{fig:grid_temperatures}.
  \item We compute the global flare temperature by integrating the flare light curve in each bandpass. The integral in each bandpass uses limits of integration equal to the start and stop times of the flare. We then divide the energy of the entire flare in the Evryscope bandpass by the energy in the TESS bandpass to obtain $R$. We then convert $R$ to $T_\mathrm{Eff}(R)$ as described above. This measures the energy associated with the average photon from the flare. The global or total flare temperature estimates the average temperature at which the flare emitted each photon. This in turn provides an estimate of the amount of UV energy associated with each photon from the flare continuum. Giving a single flare temperature representative of the entire flare duration is occasionally done when the signal to noise of the time-evolution is low (e.g. \citealt{Namekata2020}).
  \item We measure the peak temperature, defined as the mean temperature during the flare FWHM. We use the average temperature within the FWHM instead of the maximum temperature so that it will be useful for estimating UV space weather and the habitability conditions of surface life during superflare events. For example, \citet{Abrevaya2020} subject micro-organisms to likely UV conditions occurring near the peak of the 2016 Proxima superflare reported in \citet{Howard2018}. Typical fluxes during the rapid-rise and rapid-decay phases near the flare peak are usually well-described by those within the FWHM, which is why we denote this the ``peak temperature." We compute the peak temperature by randomly drawing temperatures from within the temperature uncertainties of each epoch in the flare FWHM measured in (1) and then determine the average temperature across 1000 Monte Carlo trials. The temperature near the flare peak depends on the specific flare heating conditions, so we do not assume that that the maximum (not ``peak") temperature must coincide with the epoch at the peak of the flare flux light curve. This may be too conservative of an assumption, as we find for most flares in our sample that at such low cadence the peak flux and maximum temperature do indeed coincide. This result has also been reported at 10 s cadence by \citet{Mochnacki1980}.
\end{enumerate}

Temperature measurements of epochs above 46,000 K are close to the asymptotic limit and depend strongly on small changes in the Evryscope energy measurements; we therefore flag and remove these epochs. The uncertainty of each 2 min epoch temperature measurement is estimated by error propagation. Formal errors are estimated by propagating the uncertainties in the quiescent luminosity and in the ED measurement. Systematic errors are estimated by propagating the remaining uncertainties from Section \ref{flare_energies}. Uncertainty in the global flare temperatures is estimated by integrating the total energy in each bandpass, then varying the ED and quiescent luminosity across 200 MC trials.

\subsection{Fitting model temperatures}\label{model_temperatures}

In some cases, photometric scatter during the flare obscures temperature trends from epoch to epoch. To obtain smoothly-varying model flare temperatures, we sample the energies in each bandpass from the flare fits described in Section \ref{flare_energies}. We test the model on both flares with strong signals and clear temperature trends and on flares with weak signals and noisy trends to ensure the model accurately reproduces the data as shown in Figure \ref{fig:temperature_methods}. Uncertainty in the model fits is computed by randomly drawing the ED and quiescent luminosity from their distributions of allowed values and re-fitting the model temperatures across 200 MC trials. For $\sim$10\% of flares where the temperature evolution is unclear in the epoch-by-epoch temperature measurements but clear from the model fits, we use the model temperatures in conjunction with the epoch by epoch data to determine how long each flare emitted in excess of 14,000 to 30,000 K so long as the model largely agrees with the error bars of the single epoch temperatures.

\begin{figure*}
	\centering
	{
		\includegraphics[trim= 1 1 1 1,clip, width=6.9in]{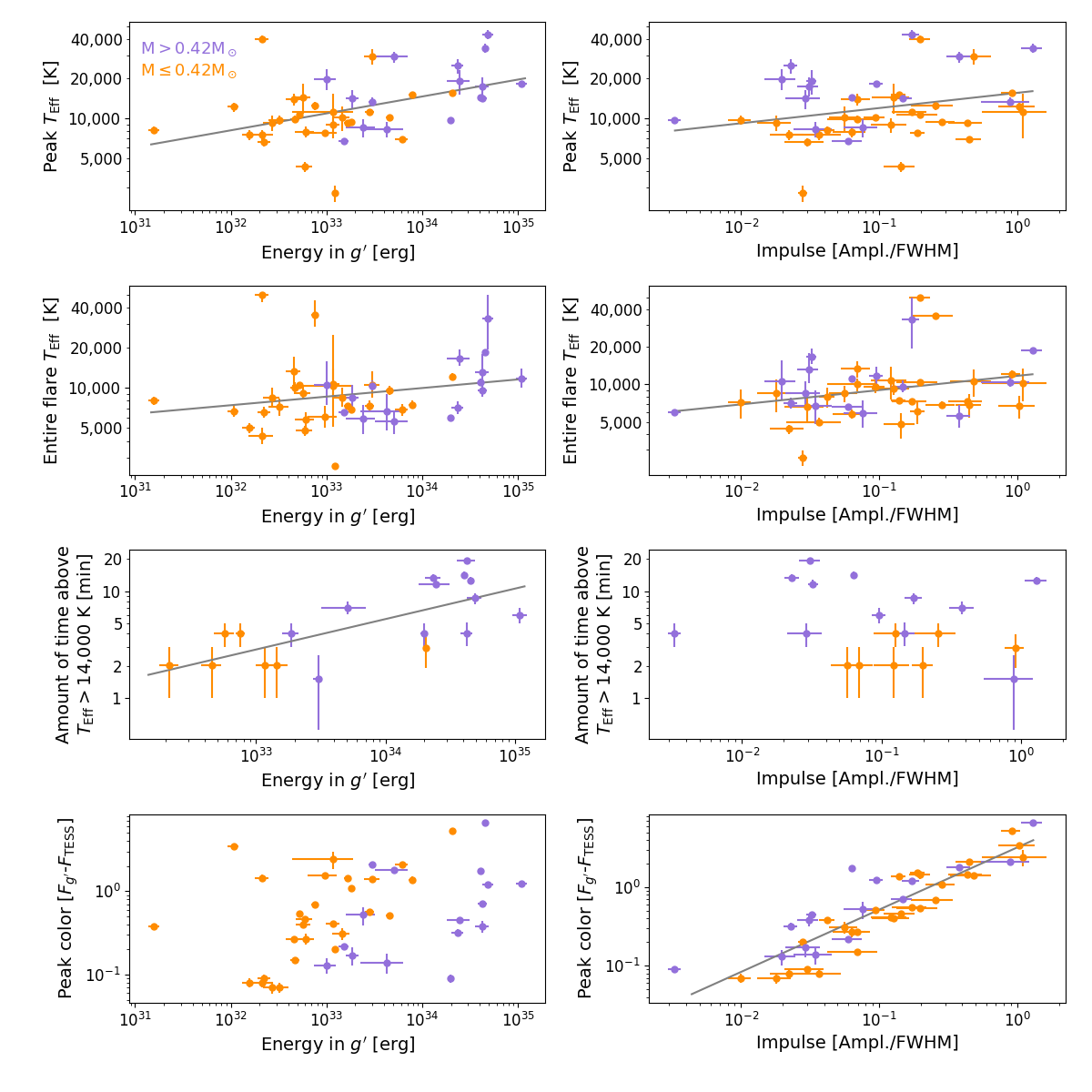}
	}
	\caption{Results: In each panel, flares are color-coded as M$>$0.42M$_\odot$ or M$\leq$0.42M$_\odot$. In each row of panels, flare energy and impulse are plotted respectively against the peak temperature ($T_\mathrm{Eff}$), total $T_\mathrm{Eff}$, amount of time a flare emits at $T_\mathrm{Eff} >$ 14,000 K, and peak flare color. Peak $T_\mathrm{Eff}$ follows a power law with $g^{\prime}$ energy. The mass gradient results in part from higher-mass stars producing higher-energy flares and in part from a selection effect for more detectable small flares from later-type stars. A weaker correlation is visible between peak $T_\mathrm{Eff}$ and impulse. Total $T_\mathrm{Eff}$ is an integrated measurement over the entire flare. Total $T_\mathrm{Eff}$ power law relationships and mass stratification are similar to those for peak $T_\mathrm{Eff}$, but with a lower y-intercept. The plots showing variation of flare energy and flare impulse with the entire ``global" flare temperature do not portray as clear a relationship as do the peak temperature plots, but are included for the sake of completeness. The amount of time a flare emits at $T_\mathrm{Eff}>$14,000 K describes higher-than-expected UV emission. A power law is visible versus energy, with a mass gradient present as before. No power law relationship with impulse is visible. Peak color is the difference in the normalized-flux amplitudes. The impulse power law is due to the relationship between flare amplitude in the $g^{\prime}$ and $T$ bands. The only free variable in the y-axis is the TESS amplitude.}
	\label{fig:grid_results}
\end{figure*}

\section{Flare energetics and morphology predict temperature}\label{results}

We find $g^{\prime}$-band flare energies that range from 10$^{31.2}$ to 10$^{35.0}$ erg and fractional-flux amplitudes in $g^{\prime}$ that range from 0.08 to 7.24. 44 large flares from 29 stars and 42 superflares from 27 stars were detected. All but two flares were from M-dwarfs, with one from a K5 and another from a K9 dwarf. The flare energy emitted in the TESS bandpass $E_T$ is related to the energy in the $g^{\prime}$ bandpass $E_{g^{\prime}}$ by log$_\mathrm{10}$ $E_{g^{\prime}}$ = 0.98 log$_\mathrm{10}$ $E_T$ + 0.62. The flare amplitude in the TESS bandpass $A_T$ is related to the amplitude in the $g^{\prime}$ bandpass $A_{g^{\prime}}$ by log$_\mathrm{10} A_{g^{\prime}}$ = 0.92 log$_\mathrm{10}$ $A_T$ + 0.86. These relationships are plotted in Figure \ref{fig:evry_tess_pwrlaws}. The flare amplitudes and energies in these relationships do not use interpolated light curves and do not suffer from the under-estimated peak flux. Only the 2 min cadence temperatures from Section \ref{tess_flares} use timestamps with interpolated flux measurements to ensure simultaneity of each 2 min epoch.

We measure the mean and 1$\sigma$ distribution of the peak $T_\mathrm{Eff}$ of our 44 flare events to be 14,000$\substack{+8,300 \\ -3,400}$ K, while the mean and 1$\sigma$ distribution of the total $T_\mathrm{Eff}$, integrated across the flare, is slightly lower at 11,000$\substack{+3,600 \\ -2,600}$ K. 43\% of the flares emit at a peak $T_\mathrm{Eff}$ above 14,000 K (the upper limit of the typical range quoted in \citet{Kowalski2013}). 23\% emit at a peak $T_\mathrm{Eff}$ above 20,000 K and 5\% emit at a peak $T_\mathrm{Eff}$ above 30,000 K. The largest and hottest flare in our sample briefly reached 42,000 K.

\subsection{Flare energetics and temperature}\label{correl_w_energy_impulse}

Flare energy correlates with $T_\mathrm{Eff}$ as shown in the left-side top and second row panels of Figure \ref{fig:grid_results}, with 10$^{35}$ erg M-dwarf flares often demonstrating twice the peak temperature of 10$^{33}$ erg flares. That larger flares are hotter is not necessarily surprising, as X-class solar flares are hotter than smaller events \citep{Feldman1996, Caspi2014}. A handful of M-dwarf superflares have also indicated a similar trend, e.g. \citealt{Robinson2005,Kowalski2010}. We find such results extend to energies of 10-1000$\times$ that of the largest solar flares and are consistent across a $\sim$10$\times$ increase in the published number of M-dwarfs with flare optical blackbody measurements at such high energies. 

While a general pattern of higher peak temperature at higher flare energy is supported on our data, the scatter in the relationship is large. For example, the 10$^{33.6}$ erg ``Hazflare" \citep{Loyd2018hazflare} had a peak temperature of 15,500 K and the 10$^{34}$ erg Great Flare of AD Leo \citep{Hawley1991} had a temperature consistent with 9000 K. Both of these events are consistent with the top left panel of Figure \ref{fig:grid_results}. Potentially more puzzling is the $\sim$10$^{31}$ erg hot flare from GJ 674 reported by \citet{Froning2019}. However, it too is consistent with our optical relation since its temperature was expected to be only 9000 K in the optical while being $\sim$40,000 K in the FUV. It is well-known that flares may emit more flux in the FUV than expected from the optical \citep{Loyd_2020}.

We note a differentiation in stellar mass in our plots. This is partly because flares from early M-dwarfs are typically larger than those from mid and late M-dwarfs \citep{Howard2019, Gunter2020} and partly because of selection effects that remove the smallest flares from the sample. We discover for the first time that a typical 10$^{35}$ erg M-dwarf flare emits above 14,000 K for $\sim$10$\times$ longer than a 10$^{33}$ erg flare, as shown in the third row left-side panel of Figure \ref{fig:grid_results}. We note some superflares never reach a temperature of 14,000 K. A clear differentiation by mass in our sample is also apparent. The differentiation seen as a function of mass is not inconsistent with the similar flare properties seen across a large range of stellar masses (e.g. \citealt{De_Luca2020, Paudel2020}). Rather, higher-energy flares occur more frequently on more massive M-dwarfs prior to spin-down (e.g. \citealt{Davenport2019, Ilin2019, Howard2019, Gunter2020}), and we find that higher-energy flares are often hotter flares. In the same span of time, more high-energy flares are likely to be observed from a young early M-dwarf than from a young mid M-dwarf star according to their respective flare frequency distributions.

\subsection{Flare impulse and temperature}\label{impulse_and_teff}
Impulse, a measure of how pronounced and rapid the flare peak is in photometry, helps to estimate when and for how long the NUV flux is greatest. \citet{Kowalski2013} defines the impulse as the fractional-flux amplitude over the FWHM in minutes, leveraging it as a proxy for the duration and intensity of flare heating at various depths in the stellar atmosphere. It is likely that photochemistry in Earth-like atmospheres may respond differently to superflares with higher impulse values \citep{Loyd2018}.

Impulse appears to correlate with flare temperature, but the power law slope is shallow as shown in the right-side top and second row panels of Figure \ref{fig:grid_results}. A relationship between impulse and $T_\mathrm{Eff}$ would indicate flare heating conditions in the stellar atmosphere are a determining factor in the blackbody properties. Flare amplitudes and FWHM values are altered by the host star's luminosity. A larger sample of temperature and impulse measurements is needed to fit power laws within each spectral sub-type to account for this effect. Binning the peak $T_\mathrm{Eff}$ into high and low-impulse sets and then randomly shuffling $T_\mathrm{Eff}$ values between bins in 10,000 MC trials, we find our observed impulse-$T_\mathrm{Eff}$ power law is reproduced by chance 1.7\% of the time for all M-dwarfs and 3.4\% of the time for M$\leq$0.42M$_\odot$ dwarfs only. However, an Anderson-Darling test finds the difference between the high- and low-impulse M$\leq$0.42M$_\odot$ flares to be statistically significant, rejecting the null hypothesis that the peak $T_\mathrm{Eff}$ of low-impulse and high-impulse events come from the same population at a p-value of 0.037. The null hypothesis is not rejected for the full M-dwarf sample.

Impulse does not clearly correlate with the time a flare emits at high temperatures. We do not find strong evidence for trends in flare temperature as a function of other variables such as stellar rotation rate. This may be due to our small sample size.

\begin{table}
\caption{Relationships between flare temperature observables and flare energy and impulse}
\begin{tabular}{p{2.9cm} p{0.9cm} p{0.9cm} p{0.9cm} p{0.9cm}}
\hline
 &  &  &  & \\
Flare Observable $O_\mathrm{fl}$ & $\alpha_E$ & $\beta_E$ & $\alpha_I$ & $\beta_I$ \\
\hline
 &  &  &  & \\
Peak T$_\mathrm{Eff}$ & 0.128 & -0.193 & 0.115 & 4.193 \\
Entire flare T$_\mathrm{Eff}$ & 0.064 & 1.811 & 0.114 & 4.07 \\
Time abv. 14,000 K & 0.285 & -8.969 & -- & -- \\
Peak color & -- & -- & 0.792 & 0.507 \\
 &  &  &  & \\
\hline
\end{tabular}
\label{table:tabulated_power_law_fits}
{\newline\newline \textbf{Notes.} We tabulate the fitted power law coefficients for each power law shown in Figure \ref{fig:grid_results}. Power laws for each flare temperature observable are given where appropriate versus flare energy and flare impulse. The power law fit for each flare observable $O_\mathrm{fl}$ versus flare energy $E_{g^{\prime}}$ is of the form log$_\mathrm{10} O_\mathrm{fl}$ = $\alpha_E$ log$_\mathrm{10}$ $E_{g^{\prime}}$ + $\beta_E$. Likewise, the power law fit for each flare observable $O_\mathrm{fl}$ versus flare impulse $I$ is of the form log$_\mathrm{10} O_\mathrm{fl}$ = $\alpha_I$ log$_\mathrm{10}$ $I$ + $\beta_I$.}
\end{table}

\subsection{Classical versus complex flares}\label{classical_music}
We classify the morphology of each flare's light curve in our sample into either ``classical" or ``complex" events. Classical flares exhibit a single flare peak while complex flares exhibit multiple peaks. Complex flares comprise the majority of the largest flares \citep{hawley2014}, making statistical comparisons of the temperatures of simple and complex flares challenging. While some flares are easily classified as having one or multiple large peaks, there is some ambiguity in the classification of other flares. While some secondary peaks significantly change the shape of the overall light curve, other flares exhibit small secondary events with an energy contribution that only perturbs the total energy (dominated by the primary peak). We include such flares in the ``classical" bin to ensure the numbers needed to assess the properties of classical versus complex superflares. For example, flare A in Figure \ref{fig:grid_evr_vs_TESS_flares} has a small secondary event that is unlikely to significantly change its total energy budget, so its light curve is best described as falling into the classical bin. Flare D, however, is best described as complex.

\begin{figure}
	\centering
	{
		\includegraphics[trim= 1 1 1 1,clip, width=3.4in]{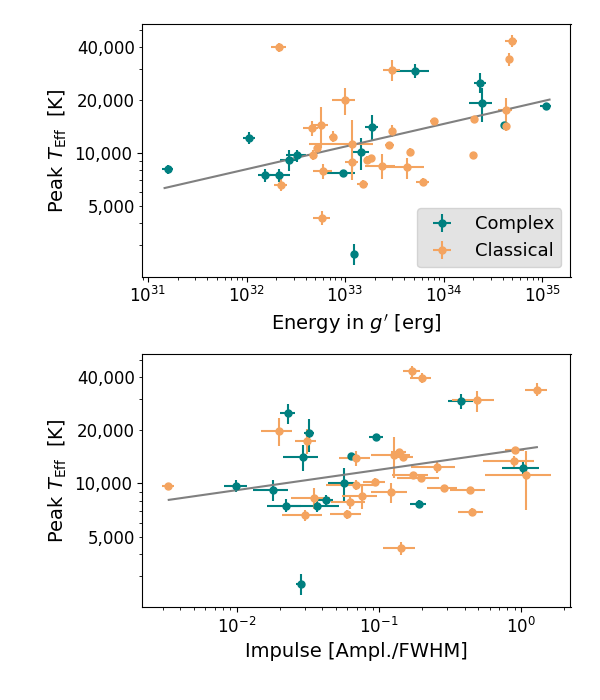}
	}
	\caption{We investigate the dependence of our peak temperature, energy, and impulse relationships on the shape of the flare. Classical flares with a single peak may correlate with temperature in a more straightforward way than complex flares do. However, we do not observe a significant difference in the energy or impulse versus temperature relationships between classical or complex flares in our (admittedly small) sample. We do note complex flares appear to have relatively lower impulse values. The appropriate trend lines from Figure \ref{fig:grid_results} are also displayed.}
	\label{fig:classical_vs_complex_fig}
\end{figure}

While the total flare energy of our sample appears to correlate with the peak flare temperature, the relationship may only hold for classical flares and not complex flares. The large amplitude and short duration of a classical flare may produce the same energy as a complex flare with a small amplitude and long duration. However, these two flares may have very different heating environments and therefore different peak temperatures. We plot the peak temperatures of the classical and complex flares in our sample against their energy and impulse in Figure \ref{fig:classical_vs_complex_fig}. We do not observe different behavior between the classical and complex flares, especially for the energy relationship. The complex flares do show lower impulse values on the same power law. We also checked these relationships for the total or ``global" flare temperatures instead of the peak temperatures and observe no difference. If this effect is physical and not a result of our small sample of superflares, it may be because secondary peaks are often of lower energy than the primary peak, acting as a perturbation to the total flare energy. Because we do not observe distinct relationships from classical and complex flares, we only fit one power law for each relationship in Table \ref{table:tabulated_power_law_fits} rather than fit classical and complex flares separately.

\begin{figure}
	\centering
    \subfigure
	{
		\includegraphics[width=3.4in]{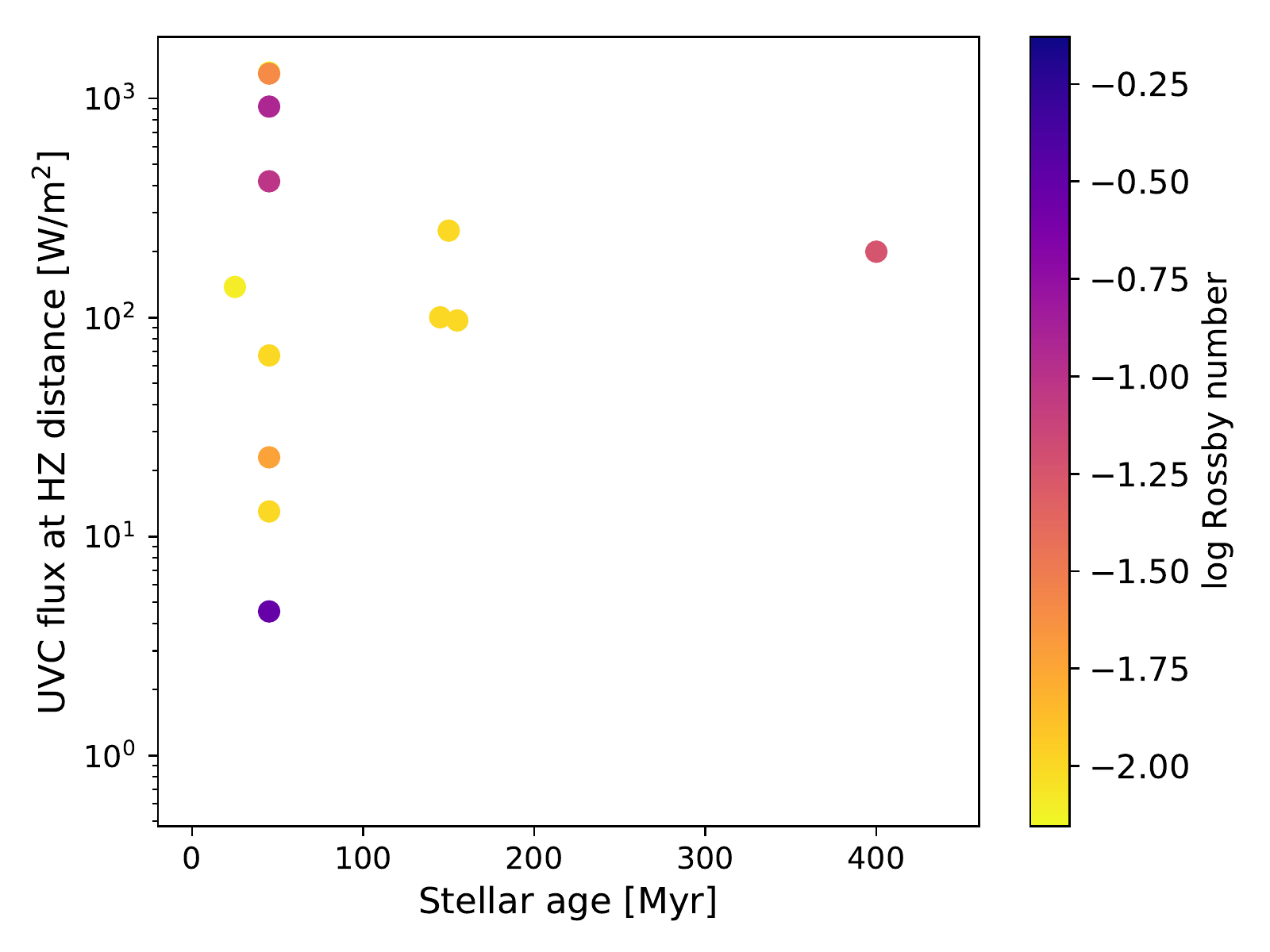}
		\label{fig:HZ_vs_mass}
	}
	\subfigure
	{
		\includegraphics[width=3.4in]{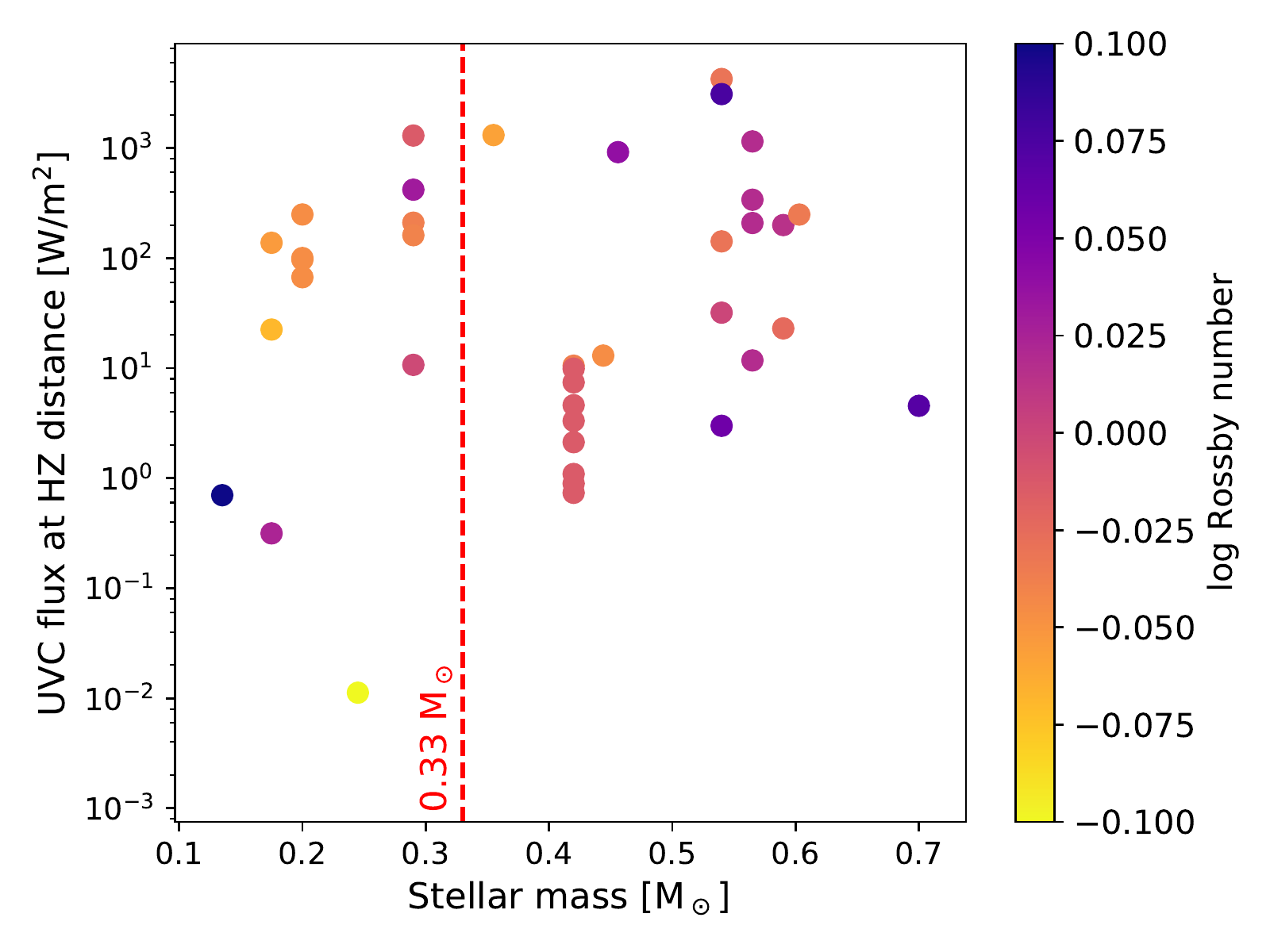}
		\label{fig:HZ_vs_age}
	}
	\caption{We plot the estimated UV-C flux emitted during the FWHM of each flare and scaled to the HZ distance from its flare star; we compare these fluxes against stellar age and mass. UV-C fluxes are estimated using the blackbody temperature of each flare. Flares are color-coded by stellar rotation rate, expressed as the Rossby number. The Rossby number is a stellar mass-independent rotation parameter useful in the large M-dwarf regime; smaller Rossby numbers correspond to faster stellar rotation periods. Top panel: Stellar ages are computed from YMG membership. The UV-C fluxes of the largest flares appear to remain approximately constant from 10$^2$ m$^{-2}$ to 10$^3$ W m$^{-2}$ for ages up to at least 200 Myr, with potentially significant consequences for young planetary atmospheres and the evolution of life during this period. Bottom panel: The UV-C fluxes at the distance of HZ planets do not appear to change significantly across the fully-convective boundary near 0.33 M$_\odot$, although a tentative but statistically-insignificant increase at higher masses may be visible in our small sample. Fully-convective stars appear to rotate more quickly.}
	\label{fig:HZ_relationships}
\end{figure}

\section{Habitability impacts of hot flares}\label{hot_flares}

The relationship between UV and optical emission is not yet well-understood \citep{Loyd2018hazflare,Loyd2018,2019AAS...23320403P, Froning2019,Loyd2020}. As a result, estimating the UV emission of stellar flares from the optical continuum may under-represent the true UV flux. For stars lacking direct UV flare observations, placing a probable lower limit on UV emission via the blackbody continuum allows us to also estimate lower limits on photo-evaporation of planetary atmospheres and constrain the conditions that might impact the evolution of surface life. The true UV emission may in fact be higher than that estimated from the optical.

\newpage

\subsection{UV-C flux in the HZ versus stellar mass}\label{versus_mass}
Larger and hotter flares are more common from more massive M-dwarfs prior to spin-down. However, the distance to the HZ is also larger for more massive stars. To determine if M-Earths orbiting lower-mass M-dwarf flare stars are more habitable, we estimate the UV-C energy of each flare using the peak blackbody temperature and the $g^{\prime}$ energy. We estimate the UV-C flux by dividing UV-C energy by the FWHM during which the energy was observed, and compute the HZ distance for each stellar mass using \citet{Kopparapu2013}. Since the fully-convective boundary occurs around $\sim$0.33 M$_\odot$ \citep{Mondrik2019}, we bin our sample of UV-C fluxes by flares from fully-convective and from earlier M-dwarfs. A slight trend towards higher UV-C fluxes at larger masses may exist in Figure \ref{fig:HZ_relationships}, but it is not clear if it is physical. We perform an Anderson-Darling test on the UV-C fluxes in each mass bin and do not find the difference to be statistically-significant (p-value $>$0.05). We therefore cannot reject the hypothesis that the UV-C space weather environment of M-Earths orbiting more massive M-dwarfs may be comparable to that for less massive M-dwarfs as estimated from the optical blackbody. A larger sample may be able to determine the relative habitability effects of hot flares as a function of mass. Because early M-dwarfs spin down faster than mid M-dwarfs, it is helpful to compare flares from stars of the same ages and different masses and different ages but similar masses. We use the Rossby number as a mass-independent relative age indicator and find our sample of flares come primarily from young stars with activity levels near the saturated regime (R$_o <$ 0.2).

\subsection{UV-C flux in the HZ versus stellar age}\label{versus_age}
Fourteen of our flares were emitted by members of young moving groups (YMG). These include 6 flares from Tuc-Hor, 4 from AB Dor, 2 from Coma Ber, 1 from Columba, and 1 from $\beta$ Pic \citep{2014AJ....147..146K, BANYAN_2014, 2017ApJ...840...87R, BANYAN_2018}, allowing us to precisely estimate their ages using Table 3 in \citet{2017ApJ...840...87R}. 16 of our flares were emitted by dwarfs that are likely to be young according to their spectral features reported in the literature but are not identifiable with any known moving group \citep{2006AJ....132..866R, 2017ApJ...840...87R}, and 13 flares come from stars with no age information available in the literature. One flare is from a $\sim$6 Gyr star (Proxima Cen) \citep{Morel2018}.

The UV-C fluxes at the HZ distance are estimated as described in Section \ref{versus_mass}. We find the median and 1$\sigma$ range of UV-C fluxes estimated at the HZ distance of flares from the YMG sample to be 120$\substack{+800 \\ -110}$ W m$^{-2}$. We find the median and 1$\sigma$ range of young stars not known to be members of YMGs to be  4.6$\substack{+137 \\ -4.3}$ W m$^{-2}$, although this sample is very small and heavily biased by one flare star with unusually-low UV-C fluxes. We find the median and 1$\sigma$ range of flares with no stellar age information to be 161$\substack{+178 \\ -158}$ W m$^{-2}$. We plot the stellar age versus UV-C flux at the HZ distance for our sample of known ages less than 500 Myr and observe similar fluxes at all ages in this range in Figure \ref{fig:HZ_relationships}. \citet{Newton2017} and \citet{Astudillo-Defru2017} find fast stellar rotators with Rossby numbers $R_o <$ 0.2 and rotation periods less than 10 d demonstrate similarly-high levels of stellar activity. Our sample is almost entirely composed of rotators with $R_o <$ 0.2, potentially explaining our consistently-high activity levels at these ages. \citet{Ilin2019} find the flare activity of cool stars decreases with increasing open cluster age, so we expect the typical flare energy (and therefore typical UV flux) will decrease at some threshold age greater than 200 Myr (not 500 Myr- only one data-point at this age). Further work is warranted.

If HZ planets orbiting $<$200 Myr stars typically receive $\sim$120 W m$^{-2}$ and often up to 10$^3$ W m$^{-2}$ during superflares, then significant photo-dissociation of planetary atmospheres may occur \citep{Ribas2016, Tilley2019}. As a point of comparison, the likely water loss of Proxima b is due to the long-term effects of a time-averaged XUV flux (including flares) of less than 1 W m$^{-2}$ \citep{Ribas2016}. The median value from the flares observed in YMGs is comparable to the $\sim$100 W m$^{-2}$ of UV-C flux estimated at the distance of Proxima b during the \citet{Howard2018} Proxima superflare. While \citet{Abrevaya2020} found 10$^{-4}$ micro-organisms would have survived the Proxima superflare, it is presently unclear what effects a 10$\times$ increase in UV-C flux would have on the evolution and survival of life prior to 200 Myr; it is possible such high rates of UV radiation could drive pre-biotic chemistry \citep{Ranjan2017,Rimmer2018}, suppress the origin of life on worlds orbiting young M-dwarfs \citep{Paudel2018}, or not impact astrobiology at all if the timescale for life to emerge is longer than 200 Myr \citep{Dodd2017, Paudel2018}.

\section{Conclusions}\label{conclude}

We have conducted the first representative survey of the optical blackbody temperature evolution of M-dwarf superflares. The multi-band photometry and analysis within our uniform sample is well-suited to statistical studies of the flare properties. We demonstrate for the first time in a large sample that flare energy and impulse are predictors of the optical temperature evolution of M-dwarf superflares. These relationships are a key step toward tailored blackbody temperatures for flares observed in photometry only, rather than having to assume 9000 K. 

Although higher-mass young M-dwarfs may emit more biologically-relevant UV flux as a consequence of frequent superflares than do lower-mass young M-dwarfs, we do not confirm that more UV-C flux from early M-dwarf superflares consistently reaches the HZ. The relative habitability of early versus mid M-dwarf planets is a topic for future work. In particular, the shorter active lifetimes of early M-dwarfs may allow planetary atmospheres to recover as the star ages via degassing \citep{Moore2020}.

In future work we will investigate enough flares observed simultaneously at 2 min cadence in order to create separate energy and impulse relationships for each spectral sub-type. We also require higher cadence to better constrain impulsive flare emission, which can occur on 10 s timescales \citep{Moffett1974}. TESS will re-observe most EvryFlare targets during Cycle 3 at 20 s cadence as part of G0 3174 to investigate the relationship between impulsive emission and temperatures of M-dwarf superflares.

\section*{Acknowledgements}\label{acknowledge}
WH would like to thank the anonymous referee for their thoughtful review of the work. WH would also like to thank Sebastian Pineda for useful comments on optical-UV flare emission.
WH acknowledges partial funding support of the Proxima Cen data and analysis through the Cycle 26 HST proposal GO 15651. WH, HC, NL, JR, and AG acknowledge funding support by the National Science Foundation CAREER grant 1555175, and the Research Corporation Scialog grants 23782 and 23822. HC is supported by the National Science Foundation Graduate Research Fellowship under Grant No. DGE-1144081. OF and DdS acknowledge support by the Spanish Ministerio de Econom\'{\i}a y Competitividad (MINECO/FEDER, UE) under grants AYA2013-47447-C3-1-P, AYA2016-76012-C3-1-P, MDM-2014-0369 of ICCUB (Unidad de Excelencia `Mar\'{\i}a de Maeztu'). The Evryscope was constructed under National Science Foundation/ATI grant AST-1407589.
\par This paper includes data collected by the TESS mission. Funding for the TESS mission is provided by the NASA Explorer Program.
\par This work has made use of data from the European Space Agency (ESA) mission {\it Gaia} (\url{https://www.cosmos.esa.int/gaia}), processed by the {\it Gaia} Data Processing and Analysis Consortium (DPAC, \url{https://www.cosmos.esa.int/web/gaia/dpac/consortium}). Funding for the DPAC has been provided by national institutions, in particular the institutions participating in the {\it Gaia} Multilateral Agreement.
\par This research made use of Astropy,\footnote{http://www.astropy.org} a community-developed core Python package for Astronomy \citep{astropy:2013, astropy:2018}, and the NumPy, SciPy, and Matplotlib Python modules \citep{numpyscipy,2020SciPy-NMeth,matplotlib}.

{\it Facilities:} \facility{CTIO:Evryscope}, \facility{TESS}

\bibliographystyle{apj}
\bibliography{paper_references}

\end{document}